# Properties of lithium aluminate for application as OSL dosimeter

A. Twardak[1], P. Bilski[1], B. Marczewska[1], J.I. Lee[2], J.L. Kim[2], W. Gieszczyk[1], A. Piaskowska[1], M. Sądel[1], D. Wróbel[1]

[1]Institute of Nuclear Physics (IFJ), Polish Academy of Sciences, 31-342 Krakow, Poland

[2]Health Physics Department, Korea Atomic Energy Research Institute (KAERI), Daejeon, 305-353, South Korea

**Abstract**

Several samples of lithium aluminate ($LiAlO_2$) were prepared in an attempt to achieve material, which can be applicable in optically stimulated luminescence (OSL) dosimetry. Both undoped and carbon or copper doped lithium aluminate samples were investigated. The results of preliminary study of theirs reproducibility, sensitivity, dose response characteristic and fading are presented. Applications in mixed field (beta, alpha, neutrons) dosimetry are discussed.

**Keywords**
Optically stimulated luminescence; lithium aluminate; $LiAlO_2$; OSL dosimetry; OSL material; neutrons measurements;

**Introduction**

Passive dosimetry is the most common measurement method in personal and environmental dosimetry. Passive dosimeters have the ability to register autonomously the absorbed doses accumulated over extended periods of time. They also feature high sensitivity, small size and independence from environmental factors such as electromagnetic or mechanical interferences. Two main techniques used in passive dosimetry are thermoluminescence (TL) and more recently optically stimulated luminescence (OSL).

OSL is a luminescence emitted from an irradiated semiconductor or insulator following exposure to light. OSL method is becoming increasingly popular precisely due to its optical nature of readout. Optical readout allows, in comparison with TL, to avoid several adverse phenomenon such as thermal quenching or unrepeatable thermal treatment. OSL is finding applications in many dosimetry areas such as personal, space, medical or retrospective and accidental dosimetry (Yukihara and McKeever 2011).

Nowadays there are known many synthetic and natural materials which exhibit OSL (Botter-Jensen et. al, 2003). The most popular and commonly used in dosimetry systems is aluminum oxide doped with carbon ($Al_2O_3$:C) (McKeever 2001). Aluminum oxide, however, has also some disadvantages for application in personal dosimetry. First of all it is not tissue equivalent ($Z_{eff}$ = 11.3) and consequently its X-ray response depends strongly on radiation energy. $Al_2O_3$:C is also not sensitive for neutrons.

The material which can be in the future considered as competitive to aluminum oxide is lithium aluminate (LiAlO$_2$). It has structure congruent to Al$_2$O$_3$, so that it can be expected to exhibit similar OSL properties. The advantage of this choice will be a small reduction of effective atomic number (10.7 for lithium aluminate) and possibility of application in neutron fields. This last feature is possible due to the fact that LiAlO$_2$ contains Li-6 isotope, which shows very high cross-section for reaction with thermal neutrons.

Preliminary results of attempts to develop a new OSL material based on lithium aluminate were demonstrated in several works. Mittani proposed doping lithium aluminate with terbium (Mittani et al., 2008), while Dhabekar and Teng suggested activation with manganese (Dhabekar et al., 2008; Teng et al., 2010). Although these attempts were not fully successful they identified opportunities and confirmed that lithium aluminate is right direction. Lee et al (2012) showed that even undoped LiAlO$_2$ may exhibit very high OSL signal. The purpose of this study is to characterize OSL properties of lithium aluminate samples and to determine influence of doping on this properties.

**Materials and methods**

For purposes of this study three sets of lithium aluminate samples were prepared at IFJ in Kraków: undoped, carbon doped and copper doped lithium aluminate. The samples were prepared using Micro Pulling Down furnace with radio frequency heating system. In all cases, the material was melted in iridium crucible. After cooling, the material from the crucible was crushed into powder. Doping with carbon was performed by the addition of graphite powder. The mass ratio of graphite to lithium aluminate was 4.7%. Copper activator was added to lithium aluminate powder as Cu$^{2+}$ ions. The concentration of Cu$^{2+}$ ions in sample was 0.1 mol%. Subsequently mixture was dried in 200 $^{\circ}$C for 3 h. For this sample 1% of lithium oxide was also added, to compensate the loss of lithium by evaporation. At the end of preparation procedure these samples were also milled into powder form.

OSL decay curves were measured using a Risø DA-20 TL/OSL reader (Risø DTU, Denmark). The reader contains 28 blue LEDs as luminescence stimulation source. The diodes emit 470 nm light and deliver approximately 70 mW·cm$^{-2}$ (at a sample surface). The detection unit is equipped in standard filter U-340, which enables to receive luminescence light in the range 300-400 nm. During experiments annealing procedure (instead of bleaching) were used to erase any possible signal. Samples were annealed in Risø reader in 400 $^{\circ}$C for 240 s. Samples were irradiated using built-in the Risø reader alpha and beta sources. For excitation with beta particles Sr-90/Y-90 (0.07 Gy/s) source was used. Alpha irradiations were performed using Am-241 source (fluence 1.7·10$^5$ cm$^{-2}$s$^{-1}$, energy at sample surface 4.5 MeV). For establishment of neutron sensitivity moderated Pu-Be (fluence 1.02·10$^6$ cm$^{-2}$s$^{-1}$) source was used. The OSL decay curves were measured in continues wave OSL mode (CW-OSL). Between irradiation and readout samples were protected from light to avoid possible signal loss. For comparison purposes Al$_2$O$_3$:C pellet or powder (manufactured by Landauer USA) were used.

**Results and discussion**

The first step in our research was to examine the reproducibility of samples dose response. The samples under study were annealed in the Risø reader, and then irradiated with beta particles with a dose of 0.7 Gy. Subsequently, CW-OSL decay curves were measured. This sequence was repeated 5 times. To evaluate the reproducibility standard deviations were calculated. The received results for all sets of samples showed good compatibility – standard deviations were below 5% (see Table 1).

Table 1. Comparison of OSL properties of investigated samples. Intensity is presented with respect to $Al_2O_3$:C.

| Sample content | Reproducibility (%) | Intensity Integral* | Initial intensity | Sensitivity ($\mu$Gy) |
| --- | --- | --- | --- | --- |
| $LiAlO_2$:C | 0.35 | 0.60 | 1.51 | 60 |
| $LiAlO_2$:Cu | 2.36 | 7.36 | 19.83 | 10 |
| $LiAlO_2$ | 4.23 | 2.89 | 7.46 | 25 |

*integral under CW-OSL decay curve from 0 to 40 s

Dose response characteristics were studied in the range from 0.3 Gy to 10 Gy. As can be seen in Figure 1 $LiAlO_2$ shows supralinearity above 1 Gy, while $Al_2O_3$:C is slightly sublinear in this range. It is noteworthy that sensitivity of undoped samples is significantly higher than $Al_2O_3$:C. Also $LiAlO_2$:Cu possesses signal, understood as integral under the CW-OSL decay curve, more than 7 times higher than aluminum oxide (see table 1). However it is important to note that adopted measurements conditions are not optimal for $Al_2O_3$:C readout. Even bigger differences can be seen when initial intensities (first point of decay curve) are compared. In this case Cu doped samples exhibit initial intensities nearly 20 times higher than aluminum oxide. These differences in proportions result from different decay rate. As can be seen in figure 2, the decay rate of lithium aluminate is significantly faster than $Al_2O_3$:C. It also can be noticed that lithium aluminate samples decay curves consist more than one component. Deconvolutions of CW-OSL decay curves show existing of at least two components – one with fast decay rate and the slow one. Due to high dose rate of used beta source (0.07 Gy/s) lower doses were not measured. To estimate sensitivity threshold standard deviation of 10 background readouts were used. The results of calculation are shown in table 1. The detection limits of investigated samples are tens of micro grays. This result suggests that lithium aluminate may be perspective material for personal dosimetry.

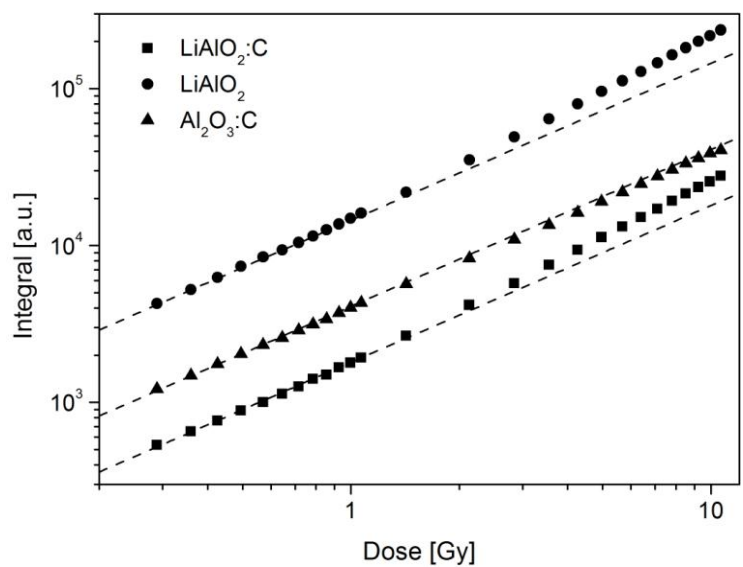

Fig. 1 Dose response characteristic of lithium aluminate samples in comparison to aluminum oxide. Broken lines indicate linearity trends.

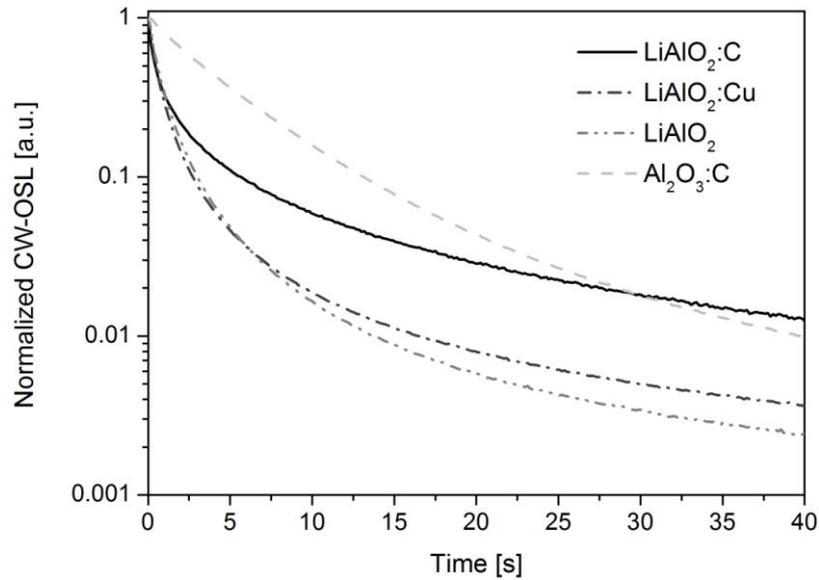

Fig. 2 Comparison of normalized decay curves of investigated samples with aluminum oxide after irradiation with beta particles. Results were normalized to range highest decay curve point.

Yukihara et al. (2011) showed that shape of CW-OSL decay curve of aluminum oxide depends strongly on the type of radiation and can be used to extracting information about LET of an unknown radiation field. Figure 3 presents CW-OSL decay curves after alpha and beta irradiation of undoped and carbon doped lithium aluminate. The undoped $LiAlO_2$ sample decay curve exhibits significantly faster decay after exposure tor alpha particles in comparison to exposition to beta particles (figure 3 a). Considerably interesting effect can be observed in case of $LiAlO_2$:C sample (figure 3 b). The relationship between CW-OSL glow curves is here opposite to undoped samples and aluminum oxide. To quantify differences between decay curves after exposure for different radiation fields the ratio R proposed by Yukihara were used. R was defined as the ratio of total CW–OSL area of beta-irradiated samples and heavy charged particles irradiated samples, after normalization of the CW–OSL curves to the initial intensity. Results of calculation of ratio R are presented in table 2. All two sets of samples except C doped set have R higher than 1. Carbon doped sample's R are lower than 1. This confirms different behavior of these samples.

Table 2. Comparison of R and N ratio of studied samples.

| Sample content | Ratio R* | Ratio N** |
|---|---|---|
| $LiAlO_2$:C | 0.82 | 7.1 |
| $LiAlO_2$:Cu | 1.61 | 3.4 |
| $LiAlO_2$ | 2.20 | 4.2 |
| $Al_2O_3$:C | - | 1.0 |

\* ratio of total CW–OSL area of beta-irradiated samples and hard charged particles irradiated samples, after normalization of the CW–OSL curves to the initial intensity

\*\* ratio of integrated signal after irradiation for neutrons to integrated signal after exposition to beta particles

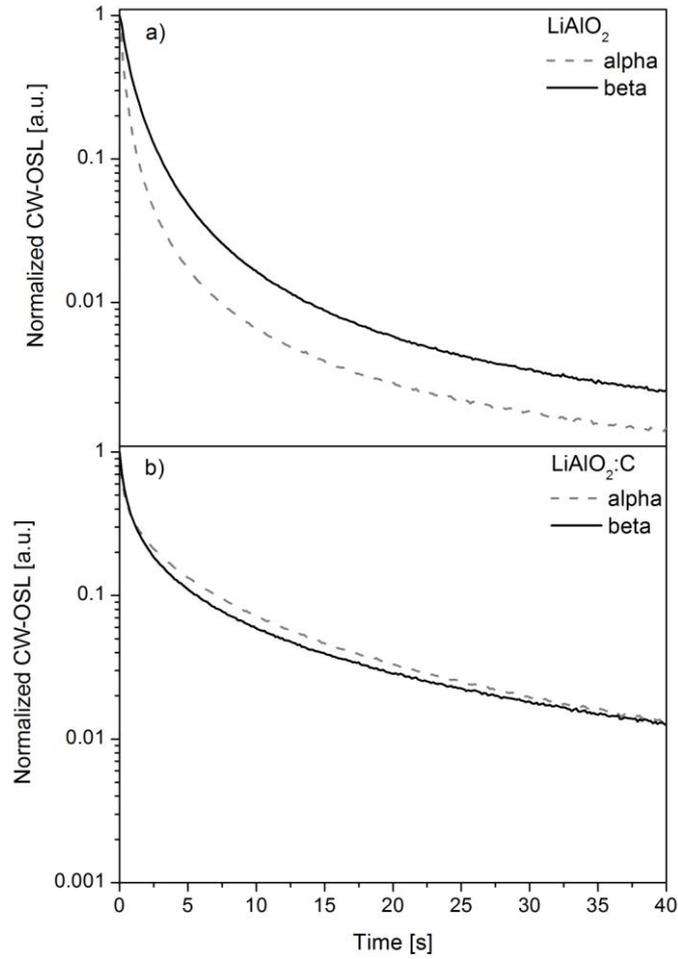

Fig. 3 OSL decay curves after exposure for alpha and beta particles. Results were normalized to range [0 ; 1]. a – LiAlO$_2$ IFJ, b – LiAlO$_2$: C

As mentioned in the introduction presence of Li-6 in lithium aluminate lattice creates opportunities for measurements in thermal neutron fields. This is due to reaction $^6$Li(n,α)$^3$H. Investigated samples were based on natural lithium, where the abundance of Li-6 isotope is about 7%. Samples were exposed to moderated neutrons from Pu-Be source (fluence 1.02·10$^6$ cm$^{-2}$s$^{-1}$) for 1 h and then CW-OSL decay curves were collected. Subsequently the response for 0.7 Gy of beta particles was measured. To quantify sensitivity for neutrons, a ratio N of integrated signal after irradiation for neutrons ($I_{neutrons}$) to integrated signal after exposition to beta particles ($I_{beta}$), were calculated:

$$N = \frac{I_{neutrons}}{I_{beta}}$$

The obtained results were normalized to Al$_2$O$_3$:C and presented in table 2. All lithium aluminate samples have ratio N significantly higher than 1. This indicates that as expected LiAlO$_2$ samples are sensitive to thermal neutron. It should be noted that Pu-Be source emits also gamma radiation and a significant part of Al$_2$O$_3$:C signal might originate due to this radiation. It should be also mentioned that received results were not corrected for fading, which is especially important in case of lithium aluminate samples. Figure 4 presents OSL fading for a post-irradiation storage of undoped and carbon doped lithium aluminate. The fading of OSL signal of undoped sample within first hour was negligible. However the signal constantly descends and fading after 10 h and 96 h was 20% and 88% respectively. Decrease of OSL signal of LiAlO$_2$:C is significant even after first hour after irradiation (55%). Nevertheless after the initial strong decline, the OSL signal considerably stabilized (fig. 4 a). The results of fading significantly improved when CW-OSL signal is integrated over the last 10 seconds of readout, however the shapes of signal decrease are similar (fig 4 b). In this case mainly slow component of OSL decay curve is integrated.

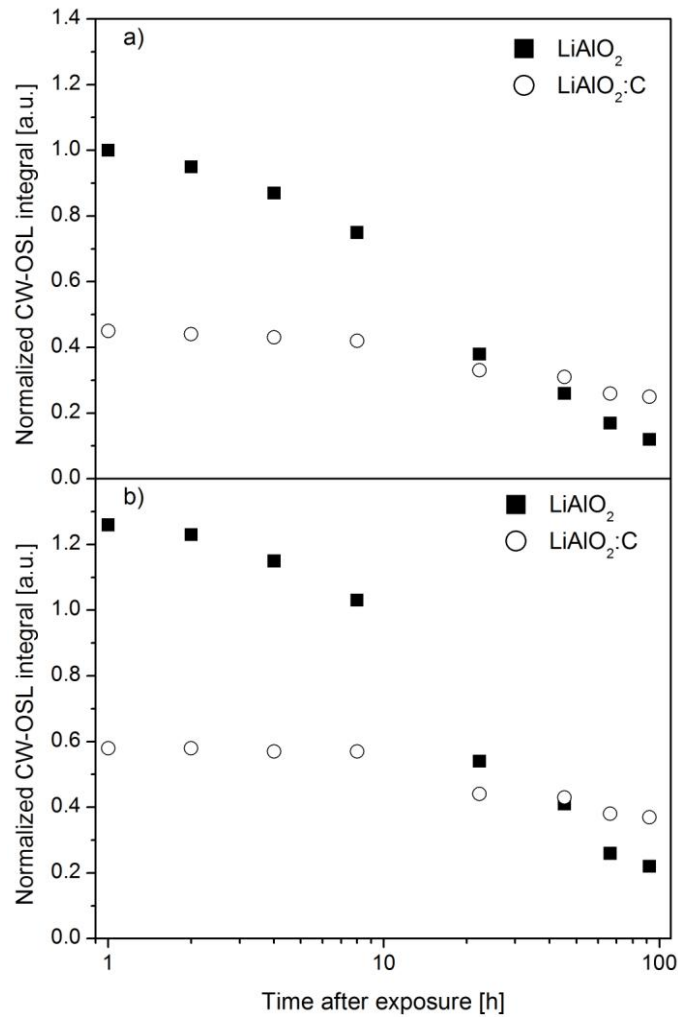

Fig. 4 Fading of undoped and carbon doped lithium aluminate. Results were normalized to signal received from readout immediately after exposure. a – signal during 40 s, b – integrated signal from last 10 s of readout (35-40s).

**Conclusions**

Three sets of lithium aluminate samples were prepared and characterized in terms of basic optically stimulated luminescence properties. All investigated of samples are sensitive for ionizing radiation and show good reproducibility. The CW-OSL decay curves shapes depend on samples' doping. The decays of all lithium aluminate samples are faster than Al$_2$O$_3$:C. The dose response is linear up to 1 Gy. Undoped and copper doped samples exhibit sensitivity several times higher than aluminum oxide, while carbon doped sample lower. The estimated sensitivity thresholds for all samples are of the order of tens of micro grays, which is good result for application in personal dosimetry. Moreover, it was shown that lithium aluminate decay curve depends on LET of radiation. What is more, LiAlO$_2$ samples are sensitive to thermal neutron. This two features make lithium aluminate an interesting material for application in mixed fields. The studied samples exhibit significant fading, but dynamics of signal loss are different for differently doped samples, what indicates the possibility of improving this characteristic by optimizing dopant composition. The presented preliminary results seems to be promising, but much work on optimization of doping and thermal treatment it is still required.

**Acknowledgments**

This work was partly supported by the National Science Centre (decision No. DEC-2012/05/B/ST5/00720) and by the Korean-Polish Science Collaboration program. Anna Twardak has been partly supported by the EU Human Capital Operation Program, Polish Project No. POKL.04.0101-00-434/08-00.